\documentclass[journal]{IEEEtran}
\IEEEoverridecommandlockouts
\usepackage{color}

\usepackage{times}
\usepackage[final]{graphicx}
\usepackage{amsmath,amsfonts}
\usepackage{amsthm}
\usepackage{amssymb,amsbsy}

\usepackage{cite}
\usepackage{multirow}

\newcommand{\ignore}[1]{}

\newcommand{\hspp}{\hspace{0.05in} }
\newcommand{\hsppp}{\hspace{0.02in} }

\newsavebox{\savepar}

\begin{document}
\title{Evolution of Physical-Layer Communications Research in the Post-5G Era}
\author{\large Vasanthan Raghavan and Junyi Li 
\thanks{The authors are with Qualcomm Flarion Technologies, Bridgewater, NJ 08807, USA.
Email: \tt{vasanthan\_raghavan@ieee.org}, {\tt junyil@qti.qualcomm.com}. }
}

\maketitle
\vspace{-10mm}

\begin{abstract}
\noindent
The evolving Fifth Generation New Radio (5G-NR) cellular standardization efforts at the
Third Generation Partnership Project (3GPP) brings into focus a number of questions on
relevant research problems in physical-layer communications for study by both academia and
industry. To address this question, we show that the peak download data rates for
both WiFi and cellular systems have been scaling {\em exponentially} with time over the last
twenty five years. While keeping up with the historic cellular trends will be possible in the
near-term with a modest bandwidth and hardware complexity expansion, even a reasonable
stretching of this road-map into the far future would require significant bandwidth
accretion, perhaps possible at the millimeter wave, sub-millimeter wave, or Terahertz (THz)
regimes. The consequent increase in focus on systems at higher carrier frequencies
necessitates a paradigm shift from the reuse of over-simplified (yet mathematically elegant)
models, often inherited from sub-$6$ GHz systems, to a more holistic view where real measurements
guide, motivate and refine the building of relevant but possibly complicated models, solution
space(s), and good solutions. To motivate the need for this shift, we illustrate how the
traditional abstraction fails to correctly estimate the delay spread of millimeter wave
wireless channels and hand blockage losses at higher carrier frequencies. We conclude this
paper with a broad set of implications for future research prospects at the physical-layer
including key use-cases, possible research policy initiatives, and structural changes needed
in telecommunications departments at universities.
\end{abstract}



\section{Introduction}
\label{sec1}
With growing demands on data rates, reduced end-to-end latencies, and connectivity
across a diversity of new applications such as the industrial Internet of Things (iIoT),
automotive, massive machine-type communications (mMTC), etc., the Fifth Generation
New Radio (5G-NR)
standard specifications for wireless systems build on prior standard releases and try
to address these complex design objectives seamlessly. The Release 15 specifications
for non-standalone and standalone deployments have been completed and approved in
December 2017 and June 2018, respectively, with continued evolution expected over the
next few months and years. Given this state of wireless evolution, 
the scope for follow-up work in terms of future releases and enhancements, and more
specifically, broader questions on relevant physical-layer research problems for study
by both academia and industry become pertinent. Such questions are important and
existential, and have been repeatedly asked in the coding theory~\cite{coding_dead},
information theory~\cite{verdu_ITdead}, digital signal processing~\cite{dsp_dead}
and communications theory~\cite{dohler_phydead} communities over many decades.

To address this question in the context of 5G-NR, we first study the
historical evolution of WiFi and cellular modems' capabilities in terms of peak download
data rates and spectral efficiencies. We show that the peak data rates have been
consistently growing {\em exponentially} over the last twenty five years for both systems,
with cellular on the brink of catching up with WiFi very soon. More importantly, the
projection of these growth rates into the near-term (next five years) can be met with
a modest increase in bandwidth acquisition and signal processing complexity of the modem.
On the other hand, a meaningful stretching of this road-map into the far future (next
ten years) would require significant bandwidth accretion at higher carrier frequencies
(than currently commercially available), commensurate signal processing complexity scaling,
and further
network densification allowing the use of higher-order modulation and coding schemes.
While the business use-cases for such increased peak data rates in future modems are
unclear as of now, applications such as pervasive health monitoring, advanced driver
assistance systems (ADAS), cellular-WiFi coexistence, etc., appear to be reasonable
drivers for sustained rate increases and latency reductions.

From a physical-layer research perspective, the evolution of 5G-NR and WiFi systems into
the millimeter wave regime signal a key transition point into the increased focus of
communications systems at higher carrier frequencies.
We show by the way of two simple examples (on delay spreads and hand blockage losses)
that the traditional abstraction of using simplified modeling techniques, sometimes
inherited from sub-$6$ GHz\footnote{Recent 3GPP standardization work has extended
the upper point of the frequency regime of interest in traditional systems from $6$ GHz
to $7.125$ GHz. Thus, the above usage should be seen with the technical caveat of
sub-$7.125$ GHz systems.} systems, for {\em systems} studies at higher carrier
frequencies is not sufficient. This is because simplified models often have no
capability to emulate unknowns that remain unknown till a deeper systemic understanding
of the impact of different components of the system on the concerned figures-of-merit.

Thus, these illustrative examples hint at a
paradigm shift in the evolution of systems research with time. In particular, systems
research has to incorporate far more of circuits and device level abstractions and
capabilities in generating relevant models, the space of possible solutions, and in
closing the feedback loop on the efficacy of the generated solutions for the intended
original problem. Both science and technology policy as well as telecommunications
departments need to evolve with these emerging trends in terms of the scope and shape
of systems research.

\section{Cellular System Road-maps}
\label{sec2}

\subsection{Historic Trends}
We start with a brief glimpse into the WiFi and cellular systems' road-maps in terms
of the log\footnote{All logarithms are to base $e$ in this work.} of the peak
download data rates and spectral efficiencies\footnote{Spectral efficiency in
bps/Hz is simplistically computed as the ratio of peak data rate and the maximum
occupied bandwidth necessary to achieve this rate.} over the past (approximate)
twenty five years. In particular, the time-period of interest is from January 1997
through December 2020 (including future forecasts) over a period of $288$ months.

For WiFi, we use open-source data
from~\cite[Table 1]{qorvo_whitepaper} on peak data rates, maximum occupied bandwidths,
and release dates (approximated to the month level) of different standard specifications.
This data is also presented in the Appendix for the sake of self-containment of this work.
For cellular systems, we use open-source data from Qualcomm (see Appendix) on the
approximate commercial sampling dates at the month level of $23$ different (in both
stand-alone and integrated forms) cellular modems as well as their corresponding
capabilities such as supported bandwidth,
peak modulation scheme, number of antennas, and signal processing
complexity~\cite{qcom_modem3}. For the signal processing complexity, we consider the
number of spreading codes in a code-division multiple access (CDMA) system, or the
number of antennas in a digital beamforming system such as Long Term Evolution (LTE),
or the number of radio frequency (RF) chains in a hybrid beamforming system such as
in 5G-NR. The modems studied in this work encompass Mobile Station Modem (MSM) 3000/3100
addressing the IS-95 (2G) specifications through the X50 modem addressing 5G-NR with
intermediate stopping points in the CDMA2000 (2.5G), WCDMA (3G) and LTE (4G) families.

\begin{figure*}[htb!]
\begin{center}
\begin{tabular}{cc}
\includegraphics[height=2.3in,width=3.2in] {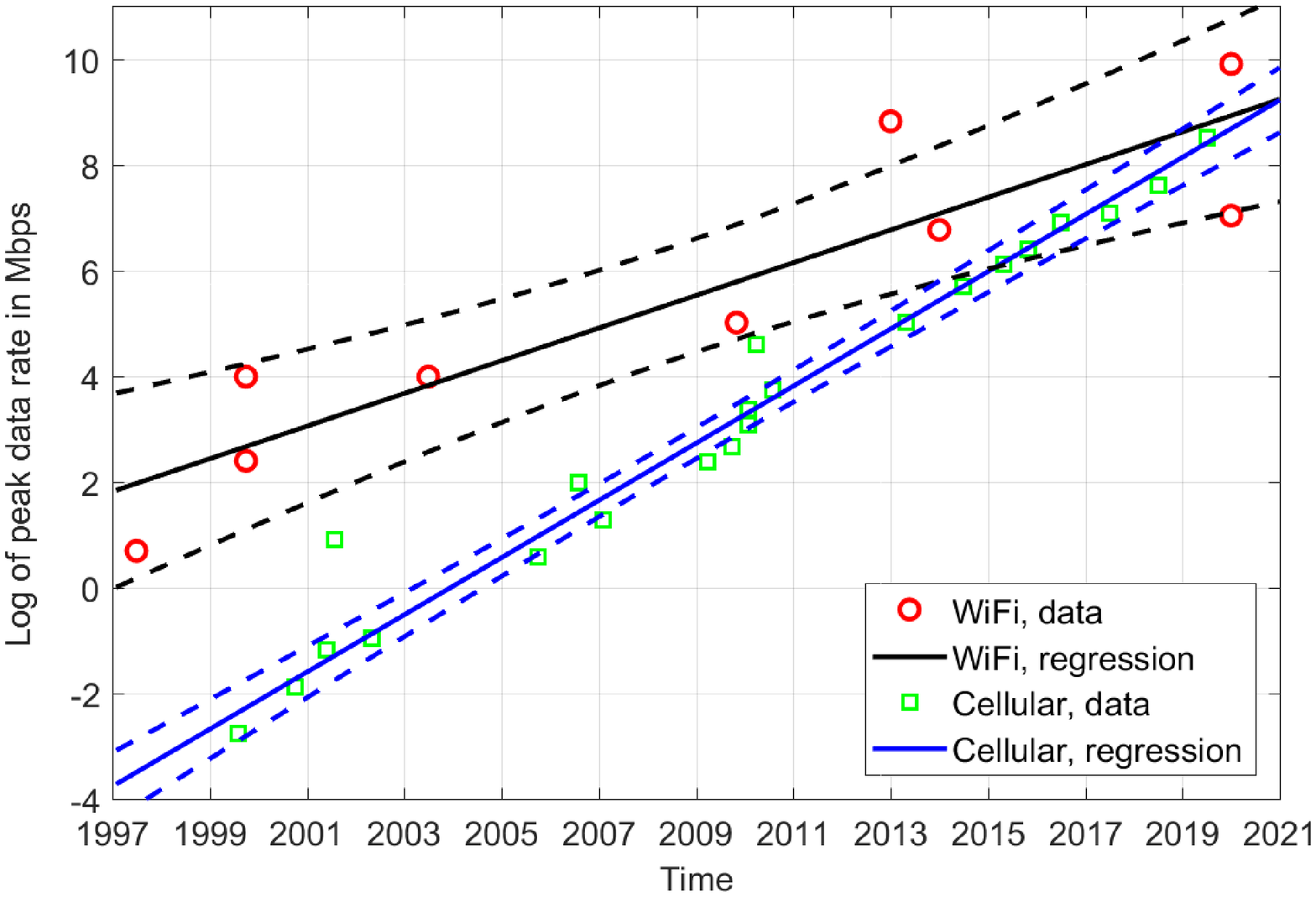}
&
\includegraphics[height=2.3in,width=3.2in]{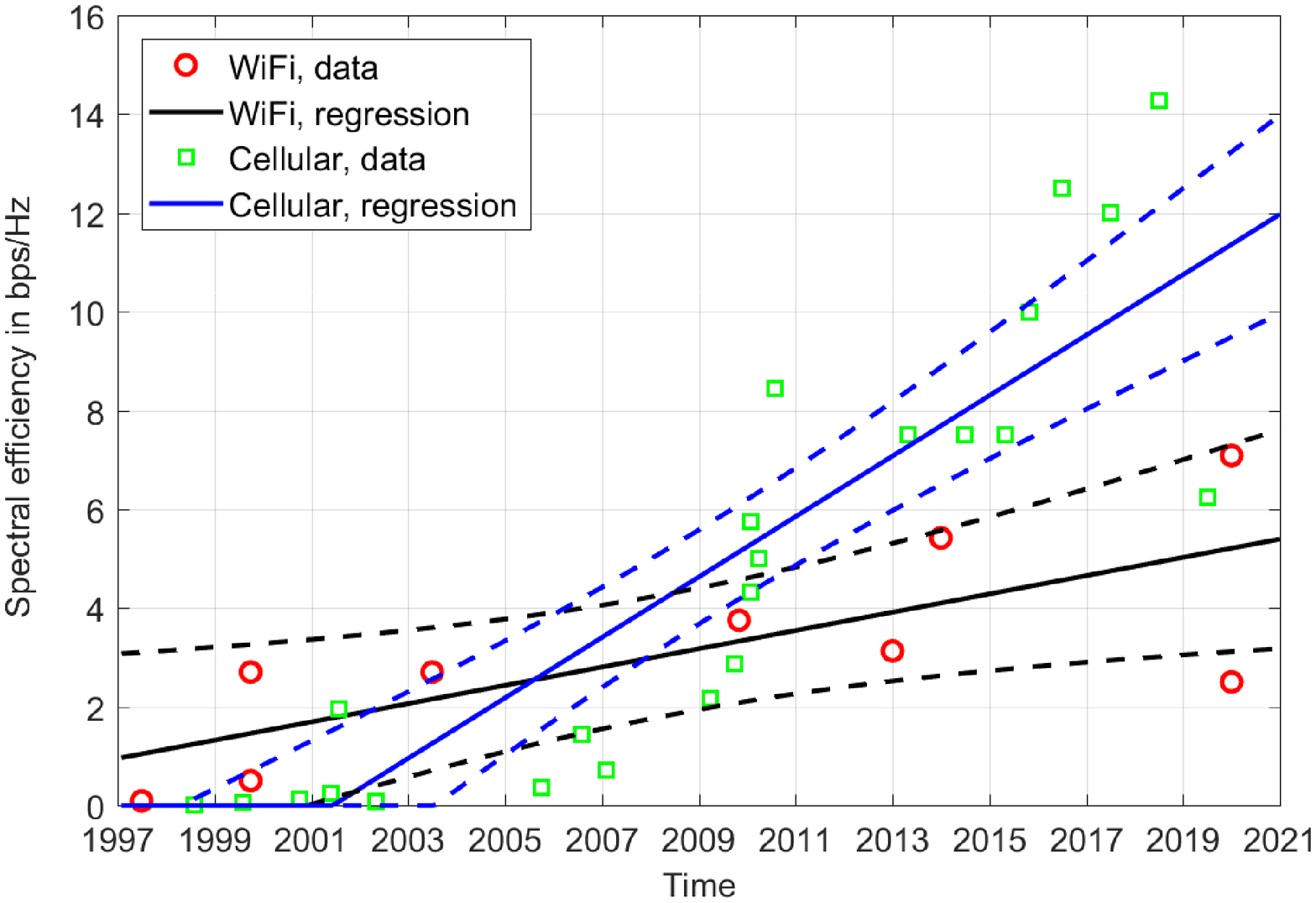}
\\
{\hspace{0.1in}} (a) & {\hspace{0.1in}} (b)
\end{tabular}
\caption{\label{fig_rate_evolution}
Evolution of (a) the log of peak data rates and (b) the spectral efficiencies
with WiFi and cellular systems, along with linear regression fits (solid lines)
and $95\%$ confidence intervals (dashed lines).}
\end{center}
\end{figure*}

The raw data for WiFi and cellular systems lead to scatter plots of
$\log \left( {\sf Peak} \hspp {\sf data} \hspp {\sf rate} \hspp {\sf in} \hspp {\sf Mbps}
\right)$ or spectral efficiency, as illustrated in Fig.~\ref{fig_rate_evolution}.
For this data, linear regression models of the form
\begin{eqnarray}
Y = \alpha_0 + \alpha_1 t + \varepsilon,
\label{eq_reg_model}
\end{eqnarray}
and a confidence interval around the regression fit are generated. The methodology behind
the linear regression modeling (including confidence interval estimation) is described
in~\cite[Chap.\ 2.4.2]{CI_linear_regression}.
In~(\ref{eq_reg_model}), $Y$ denotes the metric of interest, $t$ denotes
the month index, and $\varepsilon$ denotes the random error term. The best linear regression
fits are obtained for WiFi data (peak rates) with $\alpha_0 = 1.816$ and $\alpha_1 = 0.026$
corresponding to a standard deviation of error term of $3.60$. Similarly, for cellular,
we have $\alpha_0 = -3.764$, $\alpha_1 = 0.045$ with the standard deviation of
error term being $3.09$. Further, most of the data points (especially for cellular in the
recent past) fall within the two-sided $95\%$ confidence interval around the regression
fits, as seen in Fig.~\ref{fig_rate_evolution}. These observations suggest a reasonable
fit with the regression model in~(\ref{eq_reg_model}) for the two sets of data.

\begin{figure*}[htb!]
\begin{center}
\begin{tabular}{cc}
\includegraphics[height=2.3in,width=3.0in]{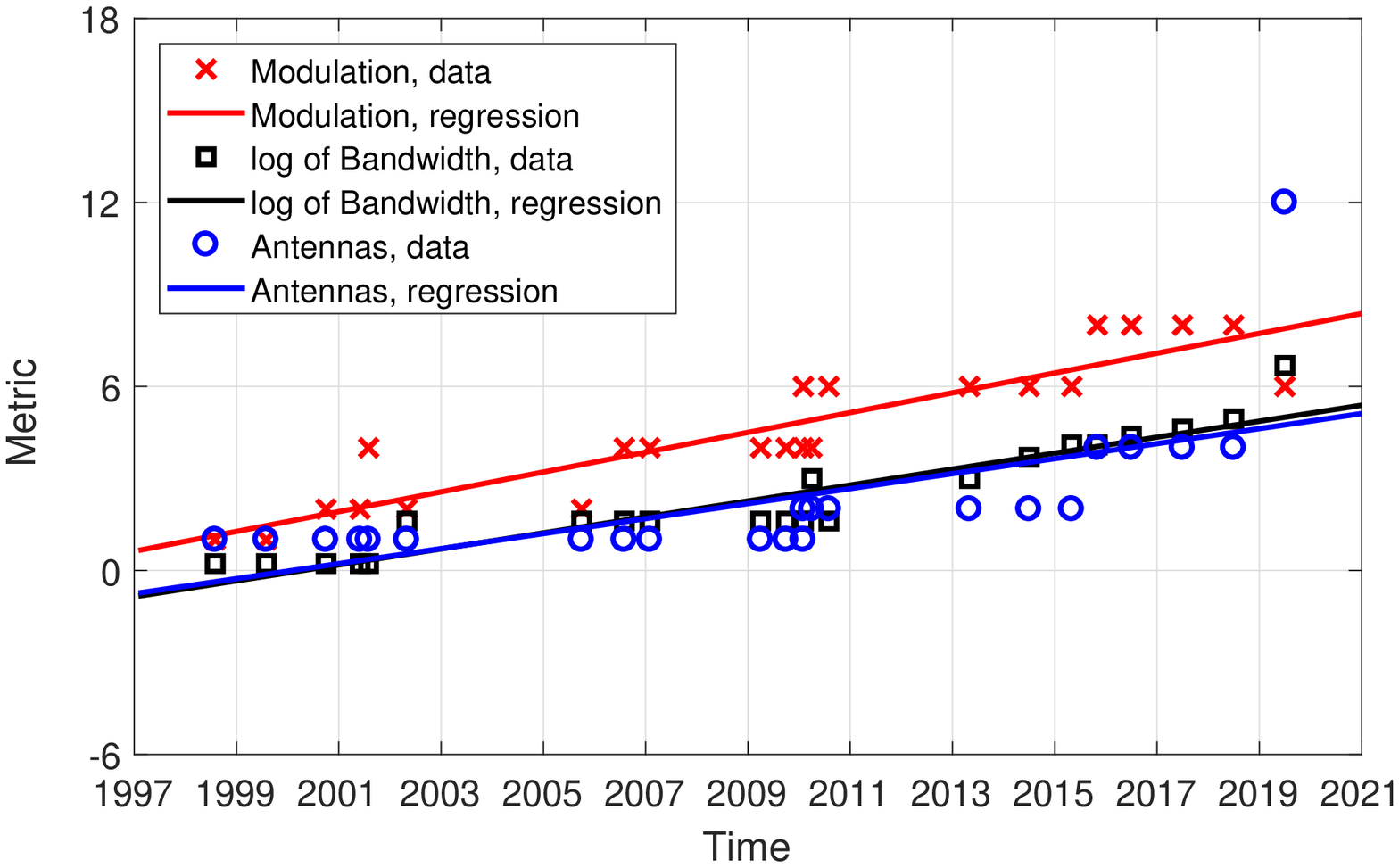}
& 
\includegraphics[height=1.8in,width=2.8in]{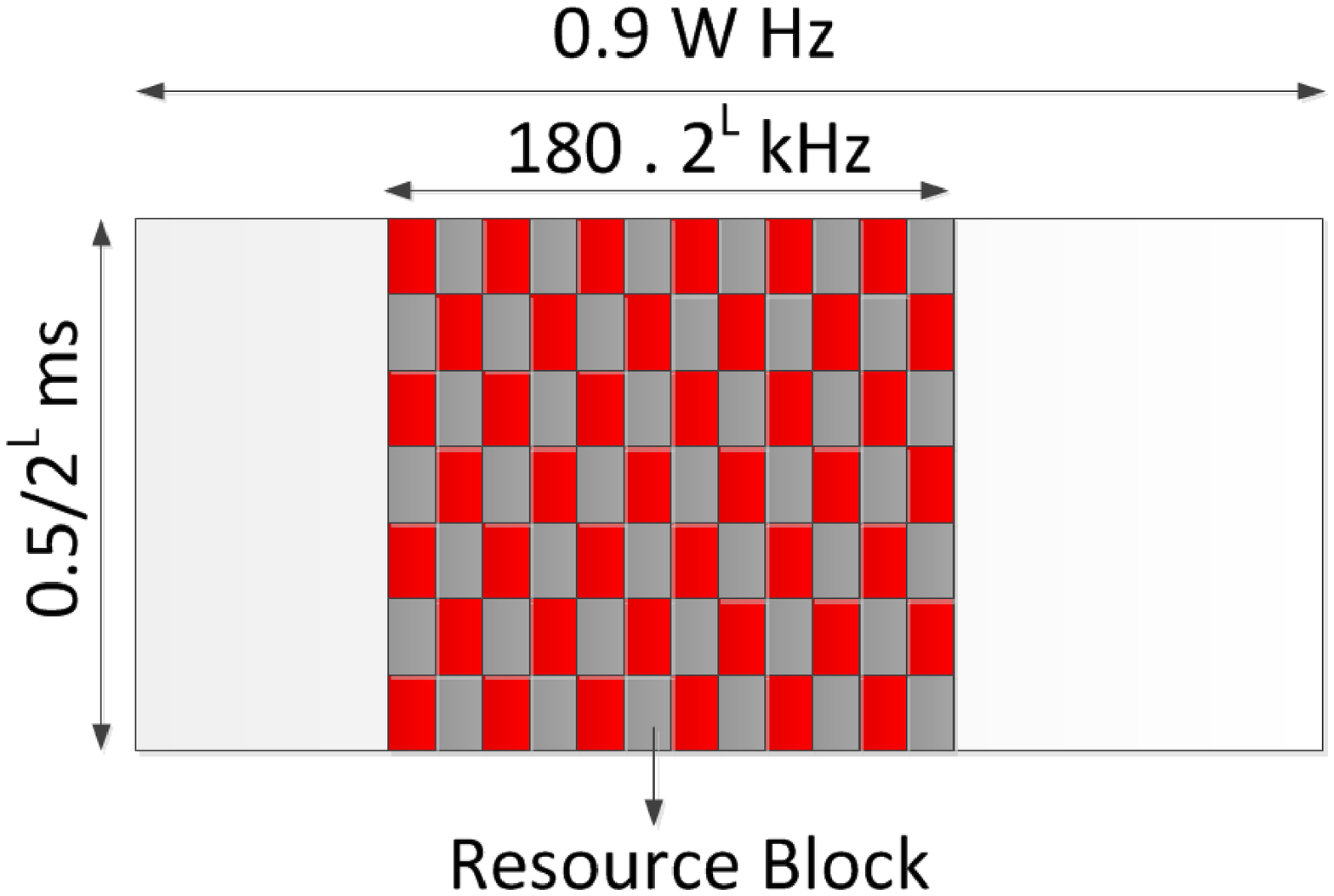}
\\
(a) & (b)
\\
\includegraphics[height=2.3in,width=3.2in]{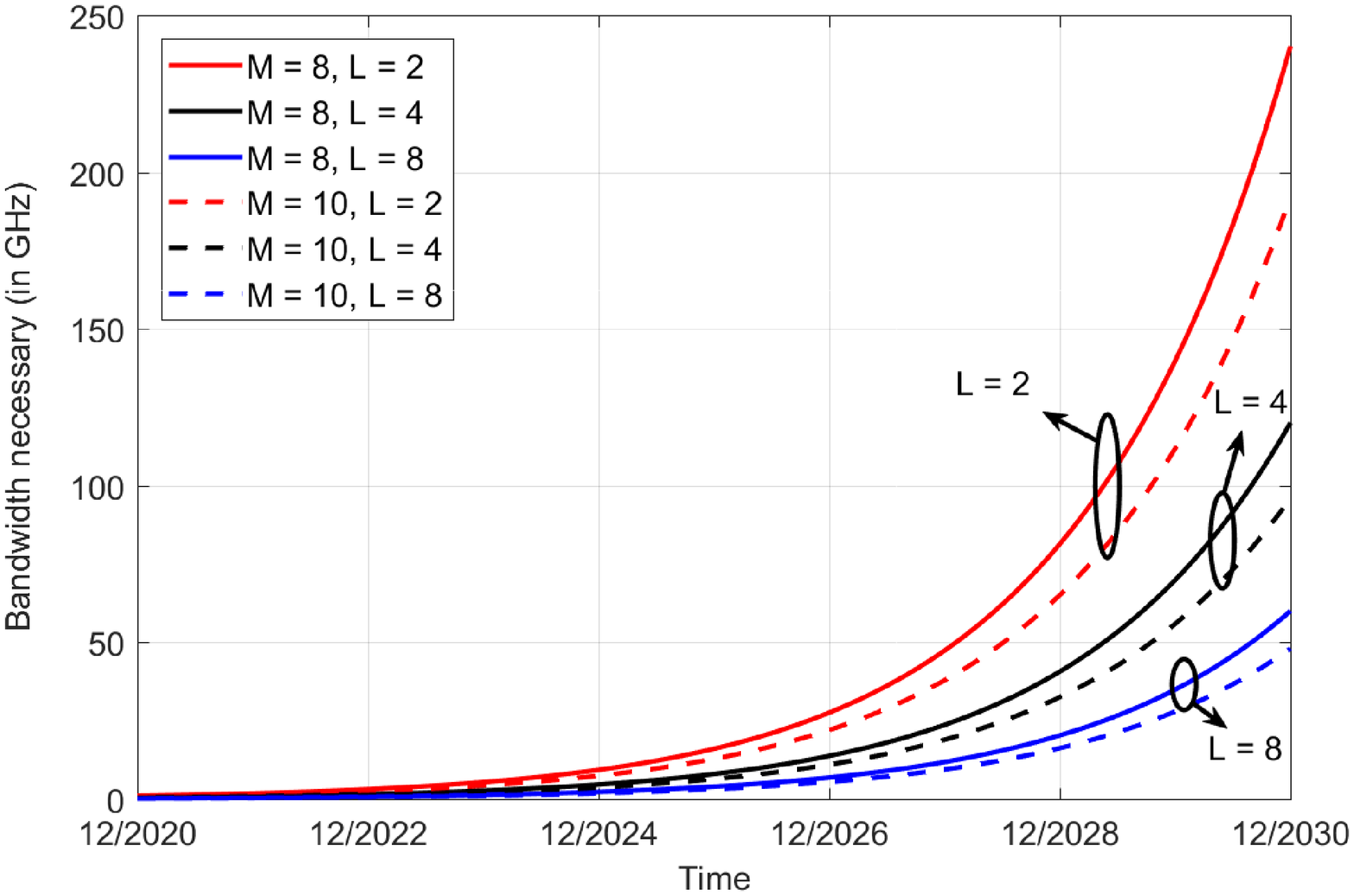}
&
\includegraphics[height=2.3in,width=3.2in]{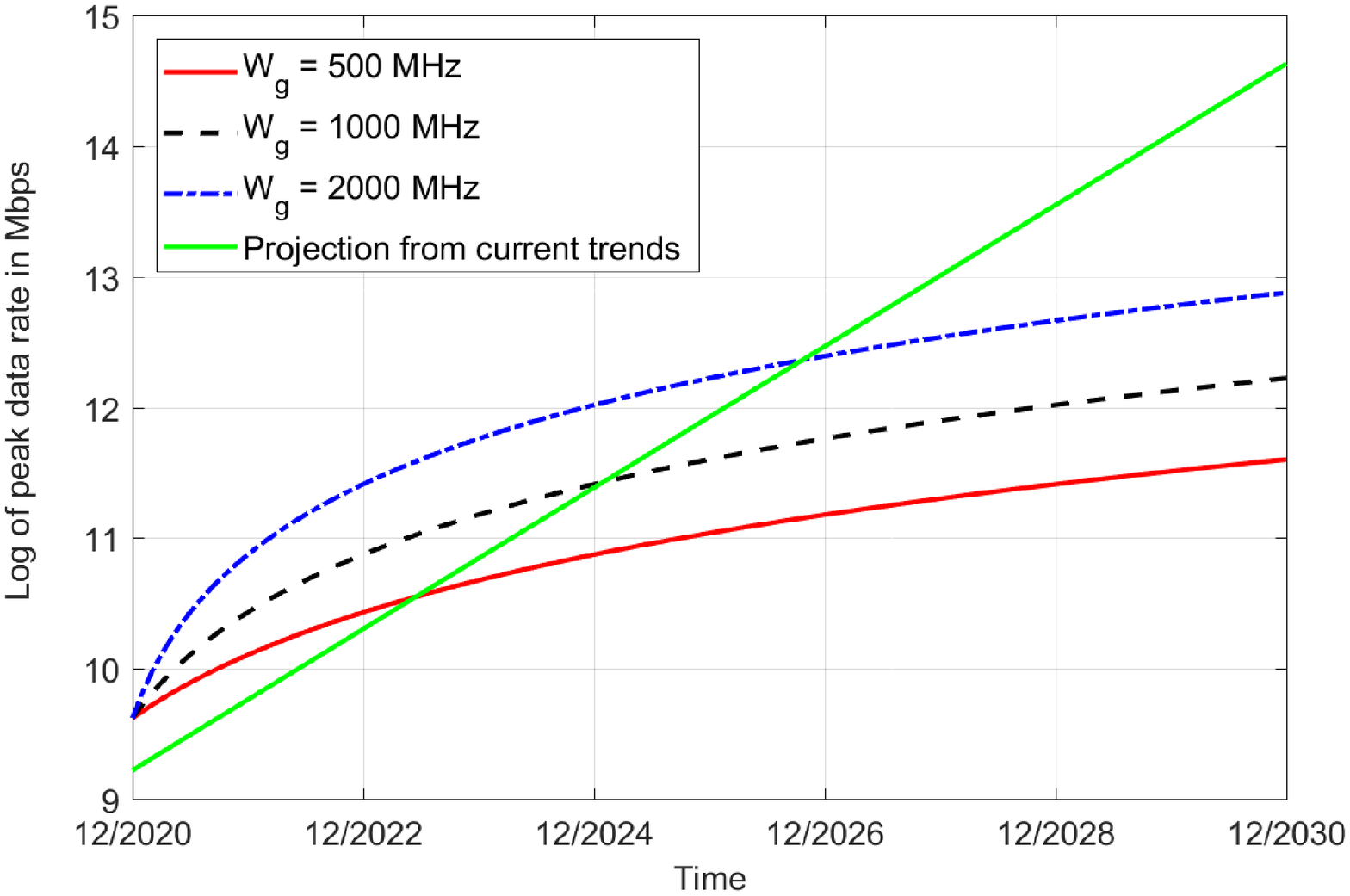}
\\
(c) & (d)
\end{tabular}
\caption{\label{fig_Wnecessary}
(a) Evolution of key factors in the exponential data growth (along with linear regression fits) \
for the cellular modems in Table~\ref{table_wifi_cellular}.
(b) Typical subframe structure assumed for a mini-slot of $7$ symbols in the
post-5G air-link specifications, building on legacy models.
(c) Bandwidth necessary to meet peak data rate demands in line with the current
trends of cellular data rate evolution. (d) Log of the peak data rate growth with
different amounts of average bandwidth accretion per year.}
\end{center}
\end{figure*}

From this study, we observe that the peak data rates have grown almost
{\em exponentially} (with time) for both WiFi and cellular systems. In particular,
the peak rates with WiFi and cellular systems have doubled over a time-period of
$\approx {\hspace{-0.03in}} 26.9$ and $\approx {\hspace{-0.03in}} 15.4$ months,
respectively. These observations suggest
that both the WiFi and cellular industries have been successful in developing
progressive road-maps with increasing capabilities for their respective modems
over time. This relentless growth has been possible due to a number of enhancements
over multiple generations of wireless standardization efforts such as:
\begin{itemize}
\item Increase in bandwidth of signaling transmissions corresponding to higher levels
of carrier aggregation~\cite{qcom_modem3}, as well as increasing bandwidth accretion
by mobile network operators over time.
\item Increase in the number of antennas and the number of RF chains/layers,
corresponding to a commensurate increase in cost, complexity, power consumption and
real-estate at the user equipment (UE) end~\cite{qcom_modem3}.
\item Densified network with smaller cell sizes and a higher frequency reuse
factor~\cite{qualcomm,wu8}.
\item Increase in total and effective isotropically radiated powers (TRPs and EIRPs).
\item Better coding schemes that can achieve higher reliabilities with lower
overheads~\cite{shrini_tom} (or higher rates). 
\item More efficient coordinated transmissions and multiple access strategies that
manage and mitigate interference, etc.~\cite{hanly_intf}.
\item Reduction in operational expenses via energy-efficient or green
transmission schemes~\cite{verdu_spec_eff,wu4,wu5,wu6,wu10}.
\end{itemize}
While such observations have been made in the past~\cite{cherry}, this work presents
evidence over a significantly longer time-frame and evolution across multiple
generations of standardization efforts.
From Table~\ref{table_wifi_cellular}, a more careful compartmentalization of some of the
specific factors leading to this exponential growth (with time) in cellular systems is
provided in Fig.~\ref{fig_Wnecessary}(a). From this plot, we note that a simple linear
regression fits the evolution of the peak modulation scheme (with bits/symbol as the
metric), log of the peak bandwidth (in MHz) and number of antennas (as a proxy for
complexity) with time. Thus, the peak bandwidth appears to be the only exponentially
scaling factor with time from the three factors studied here.

More interestingly, we also observe that while WiFi has dominated in peak rates
for almost all the time, cellular systems have been catching up rather quickly. This
trend is further clear from the spectral efficiency behavior in Fig.~\ref{fig_rate_evolution}(b),
where cellular has dominated WiFi for a long time. A number of explanations can be
offered for these observations. As carrier aggregation efforts have speeded up in
cellular systems along with coexistence in unlicensed bands, the main differentiator
in performance has been in terms of the number of RF chains/layers and power levels.
Since WiFi primarily targets indoor scenarios, the TRP/EIRP is limited to ensure
regulatory compliance, which is compensated with wider bandwidths for higher
rates, but resulting in poorer spectral efficiencies. 
Further, the lower cost factor associated with the WiFi modem (relative to the
cellular modem) limits the hardware features/capabilities and peak rates, leading to
the observed trends.

\subsection{Future Prospects for Cellular}
We now consider the implications of these trends in terms of the future
trajectory of cellular evolution. Projecting the historic trends over the next
ten years, the peak data rates should evolve as presented in the second column of
Table~\ref{table_cellular_future} (see more details later) for certain key milestone
points in time. Due to
the exponential growth rate in the past, it is not surprising to see projections
for peak data rates on the order of a few Tbps in 2030.

\begin{table*}[htb!]
\caption{Projected Cellular Data Rate Evolution According to Current Trends
and Bandwidth Necessary To Meet These Rates}
\label{table_cellular_future}
\begin{center}
\begin{tabular}{|c||c|c|c|c||c|c|c|}
\hline
& &
\multicolumn{6}{|c|}{Bandwidth necessary ($W$ in GHz) 
}
\\ 
\cline{3-8}
Time & {Peak data rates (in Gbps)}
& \multicolumn{3}{|c||}{256-QAM ($M = 8$)}
& \multicolumn{3}{|c|}{1024-QAM ($M = 10$)}
\\
\cline{3-8}
& & $L = 2$ & $L = 4$ & $L = 8$ & $L = 2$ & $L = 4$ & $L = 8$ \\ \hline \hline
Dec.\ 2020 & 
10.1 & 1.07 & 0.54 & 0.27 & 0.86 & 0.43 & 0.22  \\ \hline
Dec.\ 2022 & 
29.8 & 3.17 & 1.59 & 0.79 & 2.54 & 1.27 & 0.63  \\ \hline
Dec.\ 2025 & 
151.1 & 16.06 & 8.03 & 4.02 & 12.85 & 6.43 & 3.21 \\ \hline
Dec.\ 2030 & 
2259.9 & 240.22 & 120.11 & 60.05 & 192.17 & 96.09 & 48.04  \\ \hline
\end{tabular}
\end{center}
\end{table*}

We assume a similar subframe structure for the air-link specifications as used in
5G-NR~\cite[Sec.\ 5.3]{3gpp_CM_rel14_38802} (namely, a $15 \cdot 2^{L}$ kHz subcarrier
spacing with $L = 0, \cdots, 5$ that is possibly expanded/extended to higher carrier
frequencies), as illustrated in Fig.~\ref{fig_Wnecessary}(b). With a modest $90\%$
bandwidth occupancy, a simplistic calculation shows that the number of modulated symbols
per second (denoted as $N$) that is theoretically feasible with a bandwidth allocation
of $W$ Hz is given as
\begin{align}
& N = \underbrace{ \frac{ W 
\times 90\%}
{12 \hspp {\sf subcarriers} \times  15\cdot 2^L \cdot 10^3 } }_
{\sf 
Number \hsppp of \hsppp resource \hsppp blocks}
\nonumber \\
& {\hspace{0.3in}} \times
\underbrace{ \frac{ 12 \hspp {\sf subcarriers} \cdot 7 \hspp {\sf OFDM} \hspp
{\sf symbols} }{ 500 \hspp {\sf us}/2^L}
}_{\sf 
Resource \hsppp elements \hsppp per \hsppp second}
= 0.84\hsppp  W.
\end{align}
The peak data rate (in Gbps) assuming an $L$ layer transmission, a transmission
efficiency\footnote{Transmission efficiency is a simplistic metric to account for
overheads such as control signaling, reference and synchronization signals, and coding.
While more involved calculations as in LTE and/or 5G-NR can be performed, the
simplistic calculations done here shall suffice to extract the main trends on cellular
evolution.} of $\eta$, and a $2^M$-ary modulation scheme is given as
\begin{align}
& {\sf Peak \hspp rate} \hspp ({\sf in \hspp Gbps})
\nonumber \\
& {\hspace{0.1in}} = N \cdot \eta \cdot L \cdot M
\cdot 10^{-9}
= 0.84\hsppp W  \cdot \eta \cdot L \cdot M \cdot 10^{-9}.
\end{align}

In particular, with a practical choice such as $\eta = 70\%$ and different values
for $L$ (from $2$ to $8$) and $M$ (from $8$ for $256$-QAM to $10$ for $1024$-QAM),
the bandwidth $W$ necessary to meet the projected peak data rates (at different points
in time over the 2020-30 period) are
presented in Fig.~\ref{fig_Wnecessary}(c). This information is also presented in
Table~\ref{table_cellular_future} for certain milestone points in time over this period.
From this data, we observe that the near-term projections and demands ({\em ca.} 2023) can be
met with a spectral efficiency improvement of the cellular modem. In particular, a small
bandwidth expansion (from $800$ MHz to $2$ GHz), a hardware complexity expansion (from
$L = 2$ to $L = 4$ RF chains), and a commensurate signal processing overhead increase
are sufficient to meet these demands.

However, as we stretch out the road-map far into
the future ({\em ca.} 2030), a substantial portion of the bandwidth necessary to meet these
peak data rates (at least $45$ GHz) is not realizable except at the millimeter wave,
sub-millimeter wave and THz regimes. Further, such a rate scaling depends on licensing-related
complexities across multiple disparate geographies to be resolved before the commercialization
of these future modems. In a more realistic setting of an average year-on-year bandwidth
accretion (denoted as $W_g$) of $500$ MHz, $1$ GHz, or $2$ GHz, Fig.~\ref{fig_Wnecessary}(d)
plots the achievable peak data rates relative to the current trends projected into the
future. This plot also reinforces the ability to meet near-term trends, but not the trends
far into the future.

\subsection{Potential Use-Cases for Future Cellular Modems}
The above studies
showed that even partially meeting the historic trends on cellular data rate growth
could only be possible with increased bandwidths of signaling transmissions. Such an
increase would take us into higher carrier frequencies than those currently envisioned
or used today (e.g., $15$, $28$, $39$, $42$, $57$-$71$, or $73$ GHz, etc.).
Meeting these high data rate and low latency requirements can be extremely challenging,
if not impossible, beyond certain key milestones at price points of commercial interest.
Nevertheless, even reaching these milestones would require the design of robust, low-cost
and energy-efficient hardware (antennas, RF front-ends, etc.) that work across multiple
wide frequency bands and at higher carrier frequencies than possible commercially today.

While the necessity/use-cases for the sustained high peak data rates as
projected above are unclear as yet, three possible applications are listed below.
\begin{itemize}
\item Applications on the cellular phone can coordinate with other devices/sensors
and monitor human health near-constantly and non-invasively producing large amounts
of data. Transmitting such data from the phone to other inferencing nodes
in an {\em edge computing} framework could necessitate high data rate bursts as well as
addressing security-related concerns~\cite{wu2,wu7}.

\item Advanced driver assistance systems (ADAS)
that help with cognitive distraction
detection, collision avoidance and accident prevention, semi-autonomous driving,
etc., are expected to be a cornerstone of post-5G systems as the cellular industry
attempts to address the needs and demands of other horizontal industry segments. Such
systems are expected to consist of a number of devices/sensors performing real-time
monitoring tasks in highly dynamic environments. Coordinating such systems with
other vehicles, processing/inferencing nodes on busy streets or downtown settings,
or even other pedestrians via either the Cellular Vehicle-to-Everything (CV2X) or
the Dedicated Short Range Communications (DSRC) protocols requires high data rate
links with ultra-low latencies.

\item Coexistence of Bluetooth, WiFi and cellular systems to offer a single
{\em universal} ecosystem providing universal mobile coverage in an
{\em always on, always connected} framework has been long overdue and a possible
solution could require higher rates and lower latencies than possible today. In
this context, from Fig.~\ref{fig_rate_evolution}(a), we observe that while WiFi
has dominated in peak rates for almost all the time, cellular systems have been
catching up rather quickly. This trend is further clear from the spectral
efficiency behavior in Fig.~\ref{fig_rate_evolution}(b), where cellular has
dominated WiFi for a long time. Such a crossover is bound to have significant
impact on viable coexistence solutions.
\end{itemize}

\section{Challenges at Higher Carrier Frequencies: Illustrative Examples}
\label{sec3}
At this point, it is important to take a segue and to note that
the key progresses in communications and information theories~\cite{verdu_ITdead,dohler_phydead}
(as well as much of
technology and engineering) have been built on simple models that reflect real systems,
that are mathematically elegant, and lead to a deep intuition on system design and practice,
aptly summarized by the famous maxim of George Box~\cite{box}:
\begin{quotation}
\noindent ``Since all models are wrong the scientist cannot obtain a `correct' one by
excessive elaboration. On the contrary following William of Occam he should seek an economical
description of natural phenomena. Just as the ability to devise simple but evocative
models is the signature of the great scientist so over-elaboration and over-parameterization
is often the mark of mediocrity.''
\end{quotation}
As we march into the post-5G era of higher carrier frequencies, more care is
necessary especially since much of system design intuition relies on simplistic
models, primarily inherited from our understanding of sub-$6$ GHz systems. We
illustrate how such legacy-driven understanding can fail with 
two studies on $28$ GHz systems.

\subsection{Discrepancies in Delay Spread}
\label{sec_3a}
The delay spread is an important metric characterizing a wireless channel and is used to
understand its frequency coherence properties. The first step in estimating the delay spread
is to estimate the gains and delays of all the propagation paths from the transmitter to the
receiver. The excess and root-mean squared (RMS) delay spreads of the channel are then
computed 
as given in~\cite[(4) and (5), p.\ 6526]{vasanth_tap2018}:
\begin{eqnarray}
\tau_{\sf excess} & = & \frac{ \sum_i \tau_i p_i }{ \sum_i p_i }
\\
\tau_{\sf rms} & = & \sqrt{ \frac{ \sum_i \tau_i^2 p_i }{ \sum_i p_i} -
\left( \frac{ \sum_i \tau_i p_i }{ \sum_i p_i } \right)^2 } 
\end{eqnarray}
where $\tau_i$ and $p_i$ denote the delay and power corresponding to the $i$-th path in an
omni-directional antenna scan.

\begin{figure*}[htb!]
\begin{center}
\begin{tabular}{cc}
\includegraphics[height=2.4in,width=3.0in]{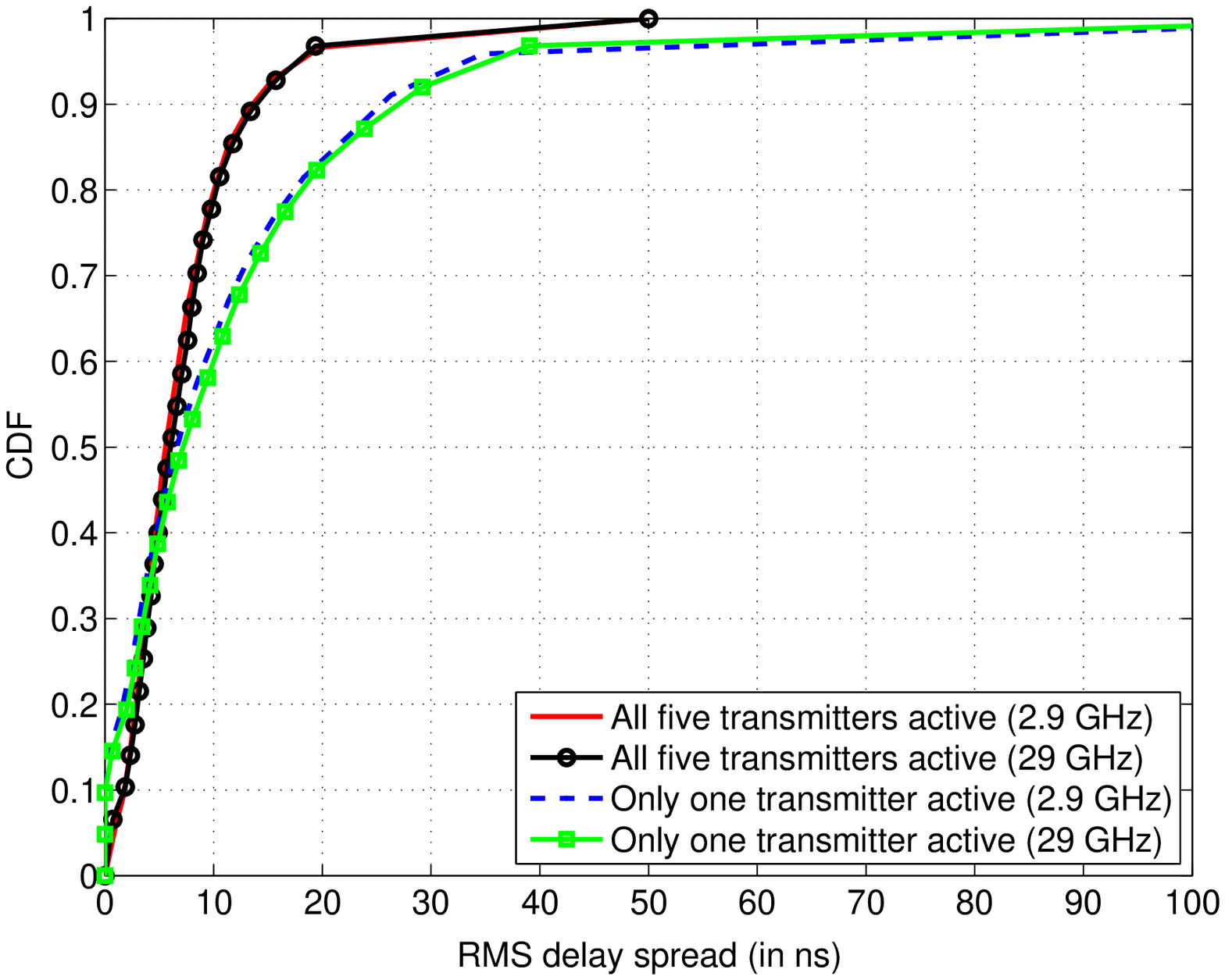}
&
\includegraphics[height=2.4in,width=3.0in]{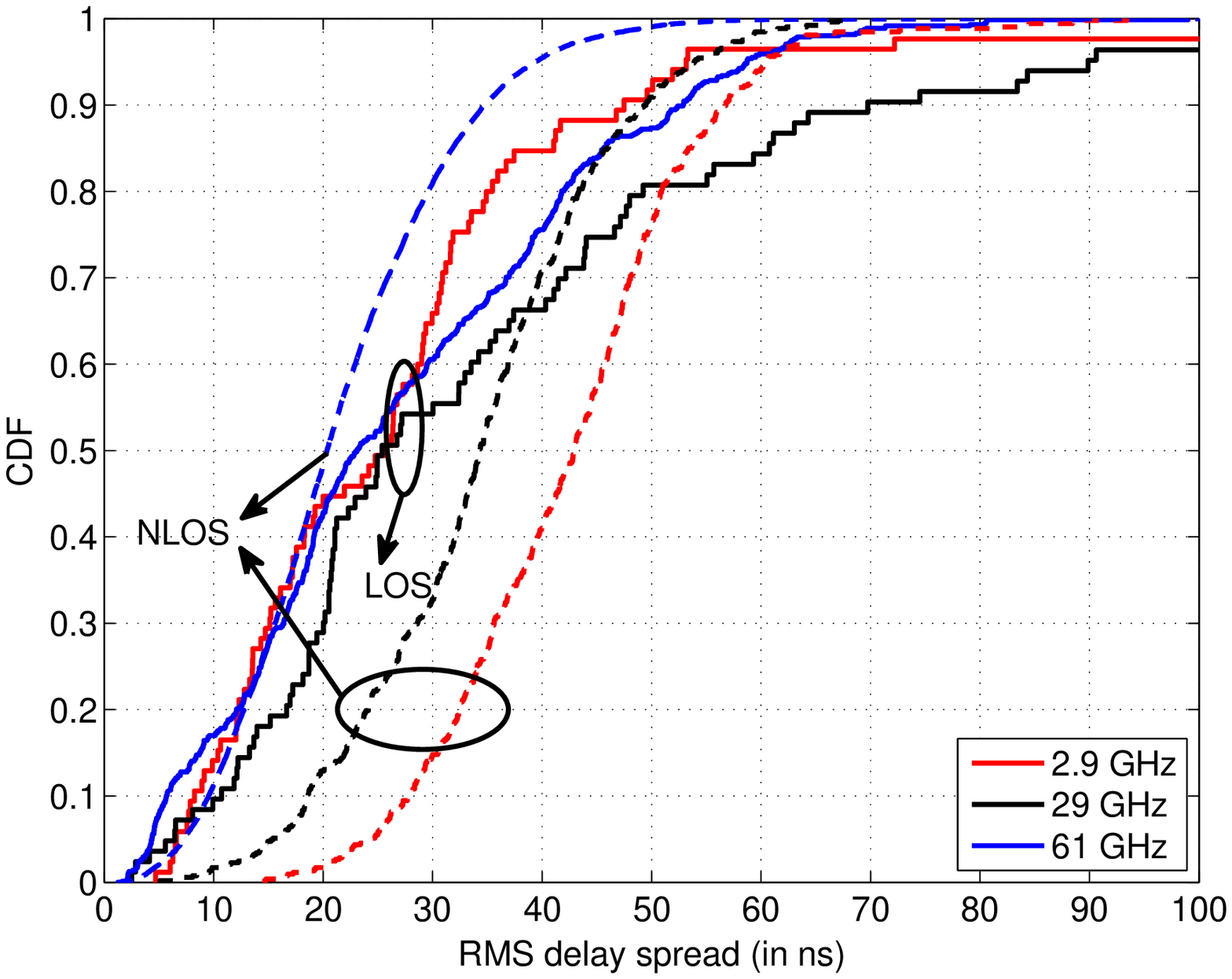}
\\
{\hspace{0.1in}} (a) & {\hspace{0.1in}} (b)
\end{tabular}
\caption{\label{fig_delay_spread}
CDF of RMS delay spreads with omni-directional antennas in a typical indoor office
deployment based on (a) ray-tracing studies and (b) measurements.}
\end{center}
\end{figure*}

In the first study, an indoor office environment (the third floor of the Qualcomm building,
Bridgewater, NJ) described in~\cite{vasanth_tap2018} and~\cite{vasanth_comm_mag_16} is
studied at $2.9$ and $29$ GHz with five transmitter locations offering coverage for the
whole area. A number of receiver locations in the building are considered and two scenarios
are studied: the transmitter that provides the best link margin is chosen for each receiver
location, and only one {\em a priori} chosen transmitter is made active for all the receiver
locations. All the transmitter and receiver locations are deployed with omni-directional
antennas at either $2.9$ or $29$ GHz. In either scenario, the gains and delays from the
transmitter to the receiver are estimated using an electromagnetic ray-tracing software
suite such as {\tt WinProp}\footnote{See more details at
{\tt https://altairhyperworks.com/product/FEKO/ WinProp-Propagation-Modeling}.}.
More details on the experiments conducted are described in~\cite{jung_gcom2015}.
Fig.~\ref{fig_delay_spread}(a) illustrates the cumulative distribution function (CDF) of
the RMS delay spread for either scenario at the two frequencies. 
From this study, we observe that transmitter diversity reduces the delay spread as expected.
More importantly, these studies show that the RMS delay spreads are comparable across $2.9$
and $29$ GHz, and the medians are less than $10 {\hspace{0.03in}} {\sf ns}$ in both cases.

In the second study, a channel sounder (described in detail in~\cite{vasanth_tap2018})
that allows omni-directional antenna scans at $2.9$, $29$ and $61$ GHz is used to study the RMS
delay spreads at the same indoor office location. The transmit-receive location pairs
used for CDF generation here are similar to those used in the ray-tracing study described
previously. Fig.~\ref{fig_delay_spread}(b) illustrates the CDF of the RMS delay spreads
for line-of-sight (LOS) and non-line-of-sight (NLOS) links at these three frequencies.
Unlike the earlier study, we observe that the RMS delay spreads of NLOS links generally
decrease with carrier frequency, whereas the LOS behavior is inconsistent with frequency.
Further, the ray-tracing study appears to significantly {\em underestimate} the true delay
spreads estimated from measurements.

Two plausible explanations are put forward in~\cite{vasanth_tap2018} to explain the
discrepancies seen with ray-tracing: i) {\em Waveguide effect} where long enclosures
such as walkways/corridors, dropped/false ceilings, etc., tend to capture more
electromagnetic energy than a simplistic LOS scenario in ray-tracing and also increase
observed delay spreads with frequency, and ii) {\em Radar cross-section effect} where
small objects of sizes commensurate with the roughness of surfaces such as walls, light
poles, metallic objects, etc., take part in propagation at higher frequencies (by
increasing the number of channel taps) and distort the delay spreads. In general,
a ray-tracing software primarily captures scattering due to buildings and large
objects/macroscopic features in the environment, and only those features that are
explicitly modeled. Thus, ray-tracing misses out on many potential (small) reflectors
and scatterers and cannot be relied on to accurately capture the delay spread in a
wireless channel at higher carrier frequencies.

\subsection{Discrepancies in Hand Blockage Loss} 
\label{sec_3b}
Another important feature of transmissions at millimeter wave, sub-millimeter wave and
THz carrier frequencies is blockage of the transmitted signal by obstructions in the
environment. In particular, electrically small
objects at microwave carrier frequencies become electrically large at higher frequencies
affecting the antenna's radiation performance. Specifically, the blockage loss associated with the
hand holding a form-factor UE has become an important metric to understand at these carrier
frequencies.

\begin{figure*}[htb!]
\begin{center}
\begin{tabular}{ccc}
\includegraphics[height=2.0in,width=1.4in] {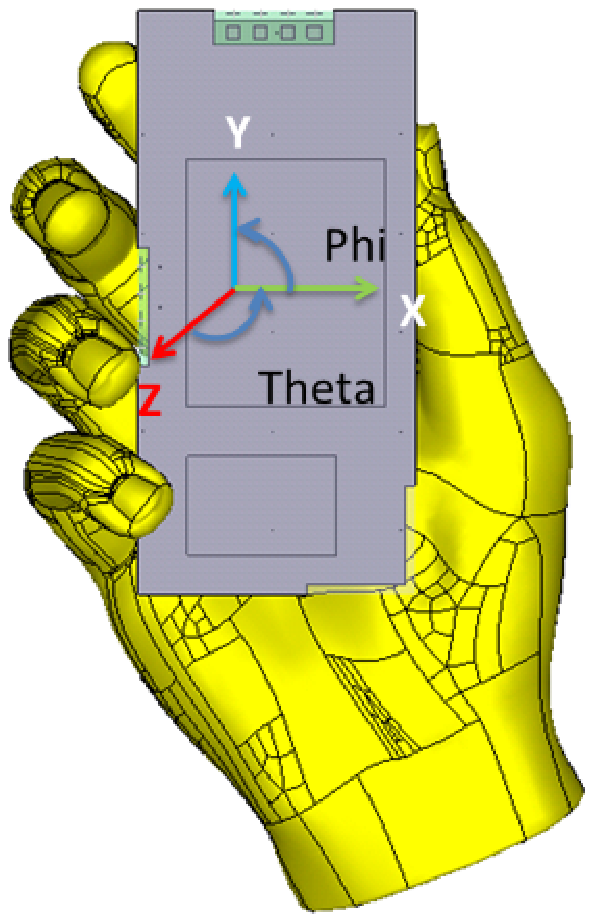}
&
\includegraphics[height=2.1in,width=2.0in] {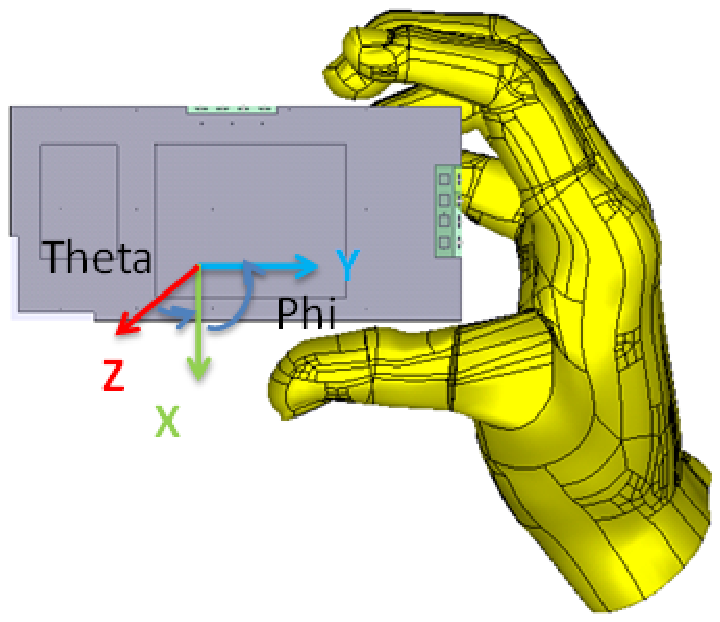}
&
{\vspace{0.1in}}
\includegraphics[height=2.3in,width=3.2in]{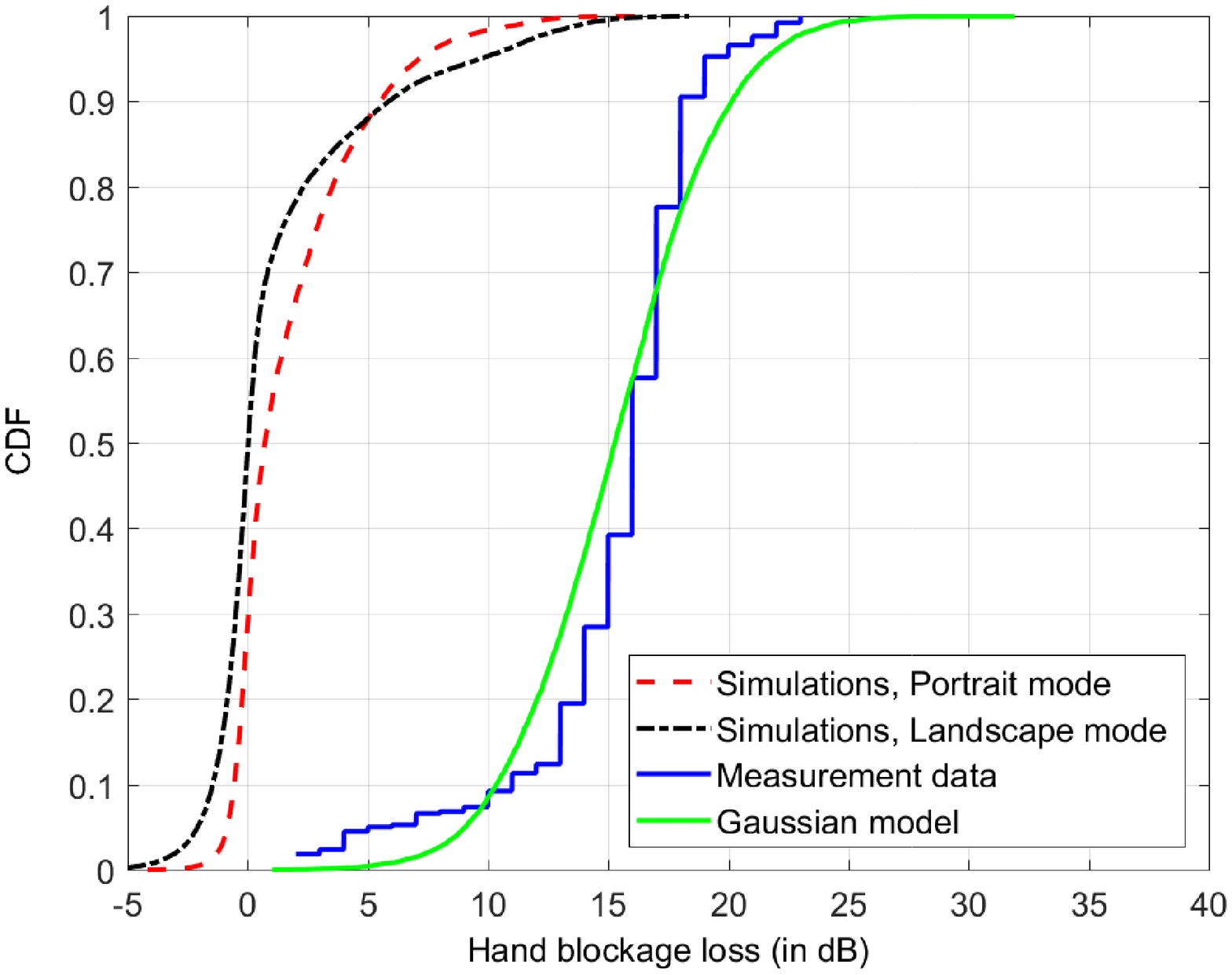}
\\
{\hspace{0.3in}} (a) & (b) & (c)
\end{tabular}
\caption{\label{fig_blockage_loss}
A typical UE design with multiple subarrays and a hand phantom model in
(a) Portrait mode and in (b) Landscape mode.
(c)
CDF of hand blockage loss with electromagnetic simulations and measurements using a $28$ GHz experimental prototype.}
\end{center}
\end{figure*}

To understand the effect of the human hand, in the first study, a simplified model of the UE
corresponding to a typical size of $60 \hspp {\sf mm} \times 130 \hspp {\sf mm}$
and designed for transmissions at $28$ GHz is studied.
As is common with sub-$6$ GHz frequencies, a simplified
model of the UE is studied in an electromagnetic simulation framework (see details
in~\cite{vasanth_blockage_tap2018} and~\cite{vasanth_comm_mag_18}).
In particular, several layers of materials emulating a realistic
form-factor design such as glass with a thickness of $1$ ${\sf mm}$, LCD shielding beneath the glass,
FR-4 board, etc., are incorporated in the simulation framework. In addition, a battery and
few shielding boxes of random sizes are placed over the printed circuit board (PCB) and are
modeled. All the metallic objects are connected to the ground plane of the PCB which covers
its bottom plane. For the antennas, multiple subarrays are placed on the long-left and top-short
edges of the UE as illustrated in Figs.~\ref{fig_blockage_loss}(a)-(b) for the Portrait and
Landscape modes, respectively. The antenna modules are designed
on a relatively low loss dielectric substrate
(Rogers 4003) and are placed on the FR-4 substrate. The antenna elements are either dipole
elements or dual-polarized patch elements with the size of each subarray being $4 \times 1$. The
antennas are designed to radiate at $28$ GHz and are simulated in Freespace (with no hand) and
with a hand phantom model, as also illustrated in Fig.~\ref{fig_blockage_loss}.

In terms of electromagnetic properties, the hand is modeled as a homogeneous dielectric with
the dielectric properties of skin tissue. These dielectric properties determine the penetration
depth of signals into the hand and the reflection of electromagnetic waves from the hand. At
$28$ GHz, a relative dielectric constant $\epsilon_{\sf r} = 16.5$ and conductivity
$\sigma = 25.8$ S/m are used in the studies~\cite{chahat}.
The UE is then simulated with and without hand using a commercial
electromagnetics simulation software suite such as {\tt CST Microwave Studio}\footnote{See
{\tt https://www.cst.com/products/cstmws} for details.}. Fig.~\ref{fig_blockage_loss}(c)
illustrates the CDF of hand blockage loss using simulated data captured as
the differential in beamforming array gains between Freespace and Portrait/Landscape
modes over a sphere around the UE.

In the second study, a $28$ GHz experimental prototype described in~\cite{vasanth_comm_mag_16}
and capturing the attributes of a 5G base-station as well as a form-factor UE design is
used to study hand blockage loss based on measurements. The UE design used in these studies
corresponds to the same setup studied with simulations earlier. In these studies (see details
in~\cite{vasanth_blockage_tap2018} and~\cite{vasanth_comm_mag_18}), the UE is
grabbed by the hand and the hand completely covers/envelops the active antenna arrays on the
long-left edge. All the subarrays at the UE side except the enveloped subarray are disabled
in terms of beam switching thus allowing us to capture the hand blockage loss in terms of
received signal strength differentials between the pre- and post-hand blocked scenarios.
Multiple experiments are performed with different
hand grabbing styles, speeds, with different air gaps between fingers, and with different
people. For each experiment, ten received signal strength indicator (RSSI) minimas spanning
the entire event from signal degradation to recovery upon removing the hand are recorded.
Link degradation is computed as the RSSI difference between the steady-state RSSI value and
the ten minimas. The empirical CDF of hand blockage loss corresponding to $38$ such experiments
is plotted in Fig.~\ref{fig_blockage_loss}(c) along with a simple Gaussian fit (specifically, of
the form ${\cal N}(\mu = 15.26 {\hspace{0.03in}} {\sf dB}, \sigma = 3.80 {\hspace{0.03in}}
{\sf dB})$) to the data.

This study illustrates the wide discrepancy between simulation-based studies and true
measurements of blockage loss. Underestimating blockage losses can lead to a poorly
designed UE with less antenna module diversity than necessary to effectuate its
seamless functioning. A number of plausible explanations can be provided for these
discrepancies. These include a poor understanding of the wide variations in material properties
(such as the human hand) at higher carrier frequencies as well as the dynamics of hand blocking,
impact of materials in the form-factor UE on signal distortion and
deterioration~\cite{vasanth_tcom2018}, capability of
simulation studies to only capture those features that can be deterministically modeled, etc.
This example illustrates the need for great care in extrapolating established techniques
for systems studies, often based on sub-$6$ GHz systems, to higher carrier frequencies.

\section{Implications on Broader Research Aspects}
\label{sec4}
The arguments put forward 
in Sections~\ref{sec2} and~\ref{sec3} focus on the importance of higher carrier
frequencies and the difficulty of simplistic simulation studies in capturing the true
impact of these systems. These observations have a number of broad implications for future
directions in physical-layer research.

\begin{enumerate}
\item
In terms of the specific blockage study of Sec.~\ref{sec3}, overcoming blockage losses at
higher frequencies requires
mechanisms that endow path diversity at far higher levels than sub-$6$ GHz systems. One
such mechanism is the use of modular UE designs with multiple antenna arrays. In
contrast to sub-$6$ GHz systems (such as LTE), such modular designs would require a careful optimization
of the antenna modules to tradeoff power consumption, diversity/spherical coverage, cost
and implementation constraints such as real-estate issues. The divergence from sub-$6$ GHz
systems in terms of UE design would require further careful
studies of antenna module placement tradeoffs~\cite{vasanth_tcom2018}. Another mechanism could be the
use of densified networks with multiple transmission points, which would also naturally
allow the use of higher-order modulation and coding schemes.

\item
At a general level,
the studies described in Sec.~\ref{sec3} clearly demonstrate the gap between traditional
simulation studies with simplistic models (often, but not always, inherited from legacy
systems) from real
observations in the field with measurements. Thus, in terms of philosophy, without closing
the gap between theory and practice of higher carrier frequency systems, the results
produced from simplified models can become meaningless in terms of the
big picture in the post-5G era.

\item
This closing of the loop requires multiple steps:
\begin{itemize}
\item A careful understanding of the different components of the system and how
they interact with each other.
\item Accurate models that capture these interactions and the contours of the solution
space along with the objective function(s) for optimization.
\item A proposed solution which can then be applied to the real scenario and
studied in terms of its efficacy in solving the original problem of interest.
\item Refinement of the model, the solution space, the proposed solution(s),
and its/their fit to the original problem.
\end{itemize}

\item
At an algorithmic level, the studies described in Sec.~\ref{sec3} suggest a possible
role for non-parametric or even machine learning-inspired
approaches~\cite{tenbrink_DL,alkhateeb_DL,wu9,wu11} in supplanting traditional statistical
signal processing and inferencing solutions in a number of applications in the cellular phone
at the sensing, processing and communications levels. However, their success would rely on
engineers' ability to extract intuition into the structure of 
these solutions.

\item
In terms of physical-layer transmissions, directional hybrid beamforming approaches over
sparse channels are of importance at higher carrier frequencies than traditional digital
beamforming
approaches~\cite{vasanth_gcom15,raghavan_jstsp,vasanth_jsac2017,vasanth_rodrigo,vasanth_tcom2018,vasanth_design,wu1}.
The cost and complexity tradeoffs in implementing such approaches
need further study as 5G standardization efforts mature and branch off to even higher
carrier frequencies.

\item
While higher carrier frequency systems are important for post-5G evolution, advances
of sub-$6$ GHz systems
cannot be ignored (or de-emphasized) in future research and development efforts. In
particular, advances in terms of form-factor UE designs with real-estate constraints
targeting lower power consumption and acceptable thermal stability, advanced physical-layer
capabilities and feature sets for different/emerging use-cases as well as highly-mobile
applications, robust coverage with carrier aggregation over multiple contiguous/non-contiguous
spectral bands, and meeting various regulatory compliance requirements (all at similar
or lower cost and complexity in implementation~\cite{wu3}) are of importance at
both sub-$6$ GHz and higher carrier frequencies. For example, while 4G systems primarily
targeted the enhanced mobile broadband (eMBB) use-case, 5G (and beyond) systems already
target other important use-cases such as ultra-reliable low latency communications (URLLC),
iIoT and mMTC with
enhancements to non-terrestrial networks (e.g., drones), integrated access and backhaul,
coexistence in unlicensed bands, positioning systems and vehicular coverage, etc.,
expected shortly. With such a diverse set of applications, both sub-$6$ GHz and higher
carrier frequency systems are expected to play a prominent role in future efforts.

\item
While fulfilling all these objectives in a reasonable manner takes a significant
amount of time and energy, such endeavors should be actively encouraged and rewarded
in terms of research funding and policy initiatives. Some recent examples in this direction
include the U.S.\ National Science Foundation's Platforms for Advanced Wireless
Research (PAWR) program ({\tt \verb|https://www.nsf.gov/funding/| \newline \verb|pgm_summ.jsp?pims_id=505316|})
and Millimeter Wave Research Coordination Network program ({\tt http://mmwrcn.ece.wisc.edu})
for fostering academia-industry interactions in the post-5G era.

\item
All this said, the fundamental dilemmas confronting a theoretician in physical-layer
research will continue to grow manifold as these systems will continue to breach
the boundaries of circuit theory, electromagnetics, communications, optimization,
statistics, signal processing, and economics. Thus, it is imperative that a modern
telecommunications department develop a core curriculum that spans these hitherto
distinct focus areas. Furthermore, it is important that such a department equip
itself with at least one advanced wireless test-bed and offer hands-on exposure to
the theory and practice of state-of-the-art telecommunications technologies to its
students and researchers.
\end{enumerate}

\section{Concluding Remarks}
\label{sec5}
The last twenty five years have been witness to a remarkable exponential scaling in cellular
modem capabilities with time. A number of technological innovations such as carrier
aggregation, higher-layer multi-antenna transmissions with more device and circuit
complexities, higher-order modulation and reliable coding schemes, network densification,
coordinated transmissions and interference management, etc., have played a key role
in this relentless growth. Sustaining these growth rates at historic levels into the
far future is both difficult as well as possibly needless due to lack of strong
business use-cases (at least as of now). Nevertheless, there are enough opportunities in
terms of both technological innovations and emergent use-cases to sustain slower growth rates in
modem capabilities with time. A central component in the evolutionary road-map of the
cellular modem would be operation over a significantly wider bandwidth across a number of higher
carrier frequencies (than currently possible today).

While such a reality is already visible today given that 5G-NR addresses millimeter
wave systems (e.g., Qualcomm's X50 modem), the focus of this work has been on the more
dramatic implications of such trends for physical-layer research
problems that would be of relevance in the next few years. This paper philosophically
argues that
{\em systems} research in its own cocoon and isolated from other areas such as
circuits/device design, electromagnetics, economics, etc., would be futile, especially
as we march inexorably to communications at higher carrier frequencies. Syncretic systems
research needs to be both encouraged and advanced from a policy standpoint, and in
the nature and scope of curriculum development and departmental structure across universities.

\section*{Acknowledgment}
The authors would like to thank Jung Ryu and Andrzej Partyka for studies on delay spread, and
M.\ Ali Tassoudji, Lida Akhoondzadeh-Asl, Joakim Hulten and Vladimir Podshivalov for studies
on hand blockage reported in this paper. The authors would also like to acknowledge the critical
feedback and encouragement of Thomas J.\ Richardson on the evolution of this paper. The authors
would like to thank the feedback from M.\ Ali Tassoudji, Kobi Ravid, Jung Ryu, Tianyang Bai,
Ashwin Sampath, Ozge H.\ Koymen, Yu-Chin Ou, Wei Yu, Erik G.\ Larsson, Emil Bj{\"o}rnson, David J.\ Love,
Srikrishna Bhashyam, Akbar M.\ Sayeed, Wei Zhang, and Durga Malladi on earlier drafts of this
article.

{\vspace{-0.05in}}
\bibliographystyle{IEEEbib}
\bibliography{newrefsx}

\appendix
\section{Explanation of Data Used in Our Study}
\label{app_data}
For the time evolution axis of WiFi and cellular data rates, we begin with Jan.\ 1997 as
Month 1 and Dec.\ 2020 as Month 288 for the data. Table~\ref{table_wifi_cellular} provides the peak
WiFi data rates and standard specification release dates. 
This table also provides the peak cellular downlink data rates, release dates and modem
capabilities used in our study.

\begin{table*}[htb!]
\caption{WiFi Data Rate Evolution 
and Cellular Data Rate Evolution with Qualcomm Modems}
\label{table_wifi_cellular}
\begin{center}
\begin{tabular}{|l|c||c|c|c|}
\hline
\multirow{9}[8]{0.5cm}{\rotatebox{90}{WiFi}} &
Standard & Specification release date
& Peak downlink data rate (in Mbps) & Bandwidth (in MHz) \\
\hline
\hline
& 802.11-1997 & June 1997 & 2 & 22 \\ \cline{2-5}
& 11a & Sept.\ 1999 & 54 & 20 \\ \cline{2-5}
& 11b & Sept.\ 1999 & 11 & 22 \\ \cline{2-5}
& 11g & June 2003 & 54 & 20 \\ \cline{2-5}
& 11n & Oct.\ 2009 & 150 & 40 \\ \cline{2-5}
& 11ac & Dec.\ 2013 & 866.7 & 160 \\ \cline{2-5}
& 11ad & Dec.\ 2012 & 6757 & 2160 \\ \cline{2-5}
& 11ax & Dec.\ 2019 & 1134 & 160 \\ \cline{2-5}
& 11ay & Dec.\ 2019 & 20000 & 8000 \\
\hline \hline
\end{tabular}
{\vskip 3mm}
\begin{tabular}{|l|c|c||c|c|c|c|c|c|}
\hline
& Modem & Approx.\ commercial & Peak downlink & Bandwidth & Modulation & Antennas & Processing
\\
& identifier & sampling date & data rate (in Mbps) & (in MHz) & & & complexity \\
\hline
\hline
\multirow{0.1}[8]{0.3cm}{\rotatebox{90}{IS-95}}
& MSM 3000 & July 1998 & 0.0144 & 1.25 & BPSK & 1 & 1 code \\ \cline{2-8}
& MSM 3100 & July 1999 & 0.064 & 1.25 & BPSK & 1 & 1 code \\ \cline{1-8}
\multirow{1}[7]{0.3cm}{\rotatebox{90}{CDMA2000}}
& MSM 5000 & Sept.\ 2000 & 0.1536 & 1.25 & QPSK & 1 & 1 code \\ \cline{2-8}
& MSM 5100 & May 2001 & 0.3072 & 1.25 & QPSK & 1 & 1 code \\ \cline{2-8}
& MSM 5500 & July 2001 & 2.4576 & 1.25 & 16QAM & 1 & 1 code \\ \cline{1-8}
\multirow{8}[8]{0.3cm}{\rotatebox{90}{WCDMA}}
& MSM 5200 & Apr.\ 2002 & 0.384 & 5 & QPSK & 1 & 1 code \\ \cline{2-8}
& MSM 6275 & Sept.\ 2005 & 1.8 & 5 & QPSK & 1 & 5 codes \\ \cline{2-8}
& MSM 6260 & Jan.\ 2007 & 3.6 & 5 & 16QAM & 1 & 5 codes \\
& & & & & QPSK & 1 & 10 codes \\ \cline{2-8}
& MSM 6280 & July 2006 & 7.2 & 5 & 16QAM & 1 & 10 codes \\ \cline{2-8}
& QSC 7230 & Mar.\ 2009 & 10.8 & 5 & 16QAM & 1 & 15 codes \\ \cline{2-8}
& MSM 7x30 & Sept.\ 2009 & 14.4 & 5 & 16QAM & 1 & 15 codes \\ \cline{2-8}
& MDM 8200 & Jan.\ 2010 & 21.6 & 5 & 64QAM & 1 & 15 codes \\ \cline{2-8}
& MDM 8200 & Jan.\ 2010 & 28.8 & 5 & 16QAM & 2 & 15 codes \\ \cline{2-8}
& MDM 8220 & July 2010 & 42.2 & 5 & 64QAM & 2 & 15 codes \\
& & & & 10 & 64QAM & 1 & 15 codes \\ \cline{1-8}
\multirow{6}[6]{0.3cm}{\rotatebox{90}{LTE}}
& MDM 9x00 
& Mar.\ 2010 & 100 & 20 & 16QAM & 2 & 2 layers \\ \cline{2-8}
& MDM 9x25/X5 & Apr.\ 2013 & 150 & 20 & 64QAM & 2 & 2 layers \\ \cline{2-8}
& MDM 9x35/X7 & June 2014 & 300 & 40 & 64QAM & 2 & 2 layers \\ \cline{2-8}
& X10 & Apr.\ 2015 & 450 & 60 & 64QAM & 2 & 2 layers \\ \cline{2-8}
& X12 & Oct.\ 2015 & 600 & 60 & 256QAM & 4 & 6 layers \\ \cline{2-8}
& X16 & June 2016 & 1000 & 80 & 256QAM & 4 & 10 layers \\ \cline{2-8}
& X20 & June 2017 & 1200 & 100 & 256QAM & 4 & 12 layers \\ \cline{2-8}
& X24 & June 2018 & 2000 & 140 & 256QAM & 4 & 20 layers \\ \cline{1-8}
\multirow{0.1}[4]{0.3cm}{\rotatebox{90}{5G}}
& X50 & June 2019 & 5000 & 800 & 64QAM & UE-specific & 2 layers \\ \cline{1-8}
\hline \hline
\end{tabular}
\end{center}
\end{table*}

\end{document}